\documentclass{ws-procs961x669}            % default, citations in superscript
\usepackage{ws-toc,ws-multind}

\begin{document}

%\titlepages

%\input preface.tex

%\cleardoublepage

%\input organizers.tex

%\blankpage                            % blank page with no running heads

%\mastertoc

%\part{First Part}{Optional Text}     % divider page, optional
%\include{proc1}
%\include{proc2}
%to switch ON running title
%\markboth{L. Hatcher}{Quantum States from Tangent Vectors}

%\wstoc{Gravitational Influence on the Quantum Speed Limit in Flavor Oscillations of Neutrino-Antineutrino System}{Abhishek Kumar Jha}
%\usepackage{amsmath,amssymb,amsthm}
%\usepackage{braket}

\title{Gravitational Influence on the Quantum Speed Limit in Flavor Oscillations of Neutrino-Antineutrino System}
%\author{A. B. Author$^*$ And C. D. Author}
%\index{author}{Author, A. B.} % or \aindx{Author, A. B.}
%\index{author}{Author, C. D.} % or \aindx{Author, C. D.}
%\email{kjabhishek@iisc.ac.in}
%\email{bm@iisc.ac.in}
%\email{mrigankad@iisc.ac.in }
%\email{mayankpathak@iisc.ac.in}
%\email{subhashish@iitj.ac.in}
\author{Abhishek Kumar Jha$^{1a}$, Banibrata Mukhopadhyay$^{1b}$, Mriganka Dutta$^{1c}$, Mayank Pathak$^{1d}$, Subhashish Banerjee$^{2e}$}
%\author{Banibrata Mukhopadhyay}
%\author{Mriganka Dutta}
%\author{Mayank Pathak}
\address{$^{1}$Department of Physics, Indian Institute of Science, Bangalore 560012, India\\
$^a$E-mail: kjabhishek@iisc.ac.in\\
$^b$E-mail: bm@iisc.ac.in\\
$^c$E-mail: mrigankad@iisc.ac.in\\
$^d$E-mail: mayankpathak@iisc.ac.in}
%\author{Subhashish Banerjee}
\address{$^{2}$Indian Institute of Technology Jodhpur, Jodhpur 342011, India\\
$^e$E-mail: subhashish@iitj.ac.in}

\begin{abstract}
We investigate the quantum speed limit (QSL) during the time evolution of neutrino- antineutrino system  
under the influence of the gravitational field of a spinning primordial black hole (PBH). We derive an analytical expression for the four-vector gravitational potential in the underlying Hermitian Dirac Hamiltonian using the Boyer-Lindquist (BL) coordinates. This gravitational potential leads to an axial vector term in the Dirac equation in curved spacetime, contributing to the effective mass matrix of the neutrino-antineutrino systems. Our findings indicate that the gravitational field, expressed in BL coordinates, significantly influences the transition probabilities in two-flavor oscillations of the neutrino-antineutrino system.
 We then apply the expression for transition probabilities between states to analyze the Bures angle, which quantifies the closeness between the initial and final states of the time-evolved flavor state. We use this concept to probe the QSL for the time evolution of the initial flavor neutrino state. %Finally, we discuss the implications of entanglement in two-flavor oscillations of the neutrino-antineutrino system in the vicinity of the PBH.
\end{abstract}

%\bodymatter

\section{Introduction}\label{Sect1}
Neutrino physics is anticipated to be crucial in addressing various astrophysical and cosmological challenges \cite{Balantekin:2013gqa}, such as neutrino-cooled accretion disks \cite{Chen:2006rra}, r-process nucleosynthesis in supernovae \cite{Surman:2008}, and leptogenesis-baryogenesis \cite{Luty:1992un},  among others. These studies require a framework that incorporates neutrinos in curved spacetime.

Neutrinos and antineutrinos are typically produced through processes such as \cite{Pontecorvo:1957cp,Bilenky:2004xm} $\beta$-decay ($n \rightarrow p + \beta^- + \bar{\nu_e}$), inverse $\beta$-decay ($p + \beta^- \rightarrow \nu_e + n$), and $\pi$-decay ($\pi^{+} \rightarrow \mu^{+} + \nu_{\mu}$), etc. There are three types of neutrinos, known as flavors: electron neutrino ($\nu_e$), muon neutrino ($\nu_\mu$), and tau neutrino ($\nu_\tau$). Their masses have not yet been precisely determined.  Any flavor state can be represented as a linear combination of three mass eigenstates, and vice-versa. The presence of non-zero masses of the mass eigenstates allows for the possible conversion between different neutrino flavor states. This quantum phenomenon is termed flavor oscillation \cite{Mohapatra:1991ng}. However, to effectively describe neutrino oscillations in curved spacetime, it is necessary to begin with the Dirac-Lagrangian formulated in curved spacetime.

In a locally flat coordinate system, gravitational interactions can be interpreted as an effective gravitational potential experienced by the system \cite{Mukhopadhyay:2007vca,Sinha:2007uh}.  
In this context, the neutrino Lagrangian density under gravity can be written as $\mathcal{L} = \mathcal{L}_f + \mathcal{L}_I$. The term $\mathcal{L}_f$ represents the free part, analogous to the Lagrangian density in flat spacetime, while $\mathcal{L}_I$ accounts for the interaction, ensuring Lorentz invariance (LI). This LI plays a crucial role in the violation of charge-parity-time reversal (CPT) symmetry in the neutrino-antineutrino system influenced by gravity \cite{Barger:1998xk,Barenboim:2002hx,Mohanty:2002et,Singh:2003sp,Ahluwalia:2004sz,Mukhopadhyay:2005gb}. This can be observed, for example, in the vicinity of rotating black holes \cite{Mukhopadhyay:2007vca,Singh:2003sp} and during the anisotropic phase of the early universe  \cite{Debnath:2005wk}. $\mathcal{L}_{I}$ introduces gravitational interactions and alters the dispersion relations for neutrinos and antineutrinos  \cite{Sinha:2007uh,Mukhopadhyay:2007vca}, a phenomenon known as the gravitational Zeeman effect (GZE) \cite{Barger:1998xk,Barenboim:2002hx,Mohanty:2002et,Singh:2003sp,Ahluwalia:2004sz,Mukhopadhyay:2005gb,Debnath:2005wk,Mukhopadhyay:2021ewd}. Previous work in this direction has examined the detailed dynamics of the neutrino-antineutrino system and its associated two-flavor oscillations \cite{Mukhopadhyay:2007vca,Sinha:2007uh}. It was shown that the effect of $\mathcal{L}_I$ is only significant in high spacetime curvature regimes and when the de Broglie wavelength of the underlying spinors is the same order as the size of the underlying gravitational source. Due to this, we choose to analyze the dynamics of our neutrino-antineutrino system around a spinning primordial black hole (PBH).

In this work, we consider a spinning PBH characterized by the Kerr metric in the Boyer-Lindquist coordinates. Based on these coordinates, we obtain an analytical expression for the four-vector gravitational potential within the Hermitian Dirac Hamiltonian. This potential introduces an axial vector term in the Dirac equation in curved spacetime. The magnitude of the gravitational vector potential is significantly affected by the angle the spinor makes with respect to the spin axis of the PBH. This gravitational vector potential term contributes to the effective Hermitian mass matrix for the neutrino-antineutrino system, which can be diagonalized through unitary transformations. We explore the unitary dynamic of this neutrino-antineutrino system using the modified mass matrix. From the viewpoint of an observer at infinity, we investigate the transition probabilities between neutrino and antineutrino in the relativistic regime for a low and a high value of specific angular momentum of the PBH. We further extend our analysis to the two-flavor oscillations of the neutrino-antineutrino system. A fundamental question emerges: how rapidly does the time evolution of the initial flavor state take place in response to changes in the specific angular momentum of the PBH? In this work, we employ the quantum speed limit (QSL) technique  \cite{Thakuria:2022taf} as the primary analytical tool to estimate the minimum time required for the neutrino-antineutrino
system to evolve from its initial flavor state to its final flavor state in the presence of a background gravitational field.

The organization of the paper is as follows: In Sec.~\ref{Sect2}, we provide the derivation of the four-vector gravitational potential in the spacetime of a spinning PBH. Sec.~\ref{Sect3} introduces the formalism of the effective mass matrix in the neutrino-antineutrino system. In Sec.~\ref{Sect4}, we discuss the oscillation of the neutrino-antineutrino system in the spacetime of the spinning PBH. Sec.~\ref{Sect5} provides a detailed analysis of the two-flavor oscillations with neutrino-antineutrino mixing in the Kerr metric. 
In Sec.~\ref{Sect6}, we investigate the quantum speed limit in two-flavor oscillations of the neutrino-antineutrino system in curved spacetime. We then conclude in Sec.~\ref{Sect7}.

\section{Gravitational potential in Kerr metric\label{Sect2} }
The Dirac Lagrangian density in curved spacetime is given by \cite{Mukhopadhyay:2007vca,Sinha:2007uh}
\begin{equation}
    {\mathcal{L}}=\sqrt{-g}~(\Bar{\Psi}i\gamma^{\mu}\overset\longleftrightarrow{D_{\mu}}~\Psi-\Bar{\Psi}m\Psi),
    \label{1.1}
\end{equation}
where the covariant derivative is,
\begin{equation}
    D_{\mu}=\partial_{\mu}-\frac{i}{4}\omega_{\nu \lambda \mu }\sigma^{\nu \lambda }.
    \label{1.2}
\end{equation}
The Lagrangian provided is applicable to both curved spacetime and locally flat spacetime, as the term $\gamma^{\mu}\overset\longleftrightarrow{D_{\mu}}$ is a Lorentz scalar. Therefore, it can be expressed as $\gamma^{d}\overset\longleftrightarrow{D_{d}}$. Here, Greek indices denote the global coordinate system, while Roman indices represent the local coordinate system. The spin connections appearing in the covariant derivative $\overset\longleftrightarrow{D_d}$ are defined as follows
\begin{equation}
    \omega_{bad} = e_{b\lambda} (\partial_d e^{\lambda}_{a} + \Gamma^{\lambda}_{\alpha\mu}e^{\alpha}_{a}e^{\mu}_{d}),
    \label{1.3}
\end{equation}
and
\begin{equation}
    \sigma^{ba}=\frac{i}{2}[\gamma^{b},\gamma^{a}].
    \label{1.4}
\end{equation}
Here $e$'s are the tetrads which connect local and global spacetimes. The tetrads satisfy the following relations: $e^{\mu}_a e^{\nu a} = g^{\mu\nu}$ , $e^{a\mu} e^{b}_{\mu} = \eta^{ab} $; where $g^{\mu\nu}$ is the metric of the global space time and $\eta ^{ab}$ is the local flat Minkowski metric. Thus, the Lagrangian can be rewritten as
\begin{equation}
   \mathcal{L}= \sqrt{-g}\bar{\Psi}(i~\gamma^d\overset\longleftrightarrow{\partial_{d}}-m+ B_d\gamma^d\gamma^5)\Psi,
    \label{1.5}
\end{equation}
with
\begin{equation}
B^d=\epsilon^{abcd}\omega_{bac},
\label{1.6}
\end{equation}
 where $B^d$ is the four-vector gravitational potential. The Lagrangian specified in Eq.~(\ref{1.5}) comprises two components: the free term and the axial vector interaction term ($B_d\gamma^d\gamma^5$). The interaction term couples with the field $B_d$\footnote{To get the covariant term $B_d$ one can write $B_d=\eta_{dc} B^c$.}, which remains constant in a local inertial frame because it originates from the background gravitational field. 

Now, we consider a spinning PBH described by the Kerr metric. The Kerr metric in the Boyer-Lindquist coordinates is \cite{Misner:1973prb}
\begin{align}
    ds^2 &=(1-2Mr/\rho^2)dt^2+(4Mar\sin^2\theta/\rho^2)dt d\phi-(\rho^2/\Delta)dr^2 \nonumber\\
    & -\rho^2 d\theta^2-\left[(r^2+a^2)\sin^2\theta+(2Ma^2r\sin^4\theta/\rho^2)\right]d\phi^2,
    \label{1.7}
\end{align}
where $\rho^2=r^2+a^2 \cos^2\theta$, $\Delta=r^2-2Mr+a^2$, $M$ is the mass of the black hole, $a$ is the angular momentum per unit mass ($-M\leq a\leq M$), and $\theta$ is the angle the particle makes with respect to the spin-axis of the black hole. Here, we choose $M=c=\hbar=1$. Hence, the event horizon (“one-way membrane”) of the rotating black hole  is 
\begin{equation}
    r_{+}=1 + \sqrt{1 - a^2}.
    \label{1.8}
\end{equation}
For $a=0$, Eq.~(\ref{1.7}) reduces to the Schwarzschild metric.

 Without loss of generality, we choose the Schwinger gauge of tetrads \cite{Ghosh:2020lhw,Schwinger:1963zz,Gorbatenko:2011ui,Neznamov:2018mer}, given by
\begin{align}
    e^t_0 & =\sqrt{g^{tt}},\hspace{0.5em} e^r_1=\sqrt{\Delta}/\rho,\hspace{0.5em}e^\theta_2=1/\rho,\nonumber\\
    e^\phi_3 &=1/(\sin\theta \sqrt{\Delta}\sqrt{g^{tt}}),\hspace{0.5em}e^\phi_0=2Mar/(\rho^2 \Delta \sqrt{g^{tt}}).
    \label{1.9}
\end{align}

\def\figsubcap#1{\par\noindent\centering\footnotesize(#1)}
\begin{figure}[t]%
\begin{center}
  \parbox{2.1in}{\includegraphics[width=2.2in]{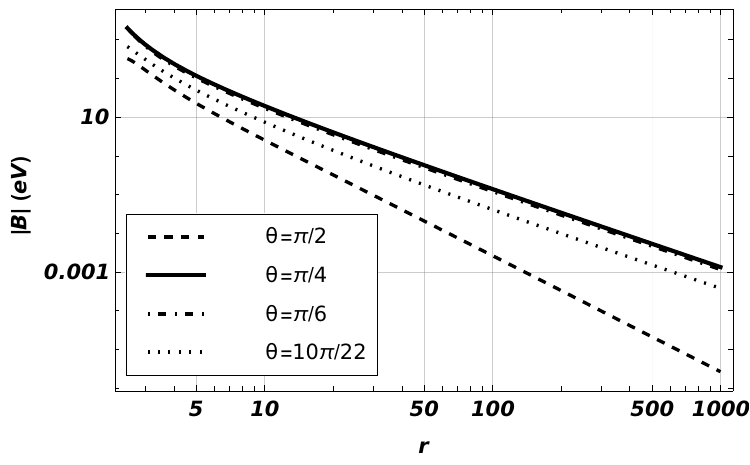}
  \figsubcap{a}}
  \hspace*{4pt}
  \parbox{2.1in}{\includegraphics[width=2.2in]{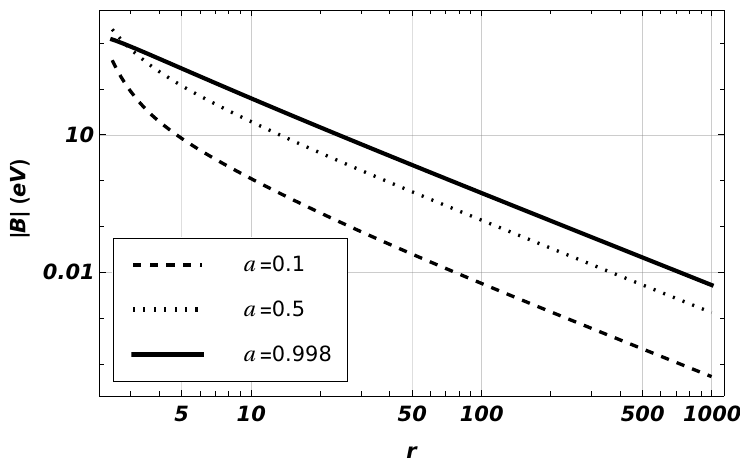}
  \figsubcap{b}}
  \caption{The magnitude of gravitational vector potential $|\textbf{B}|$ as a function of the radial distance $r$ from the black hole (a) at four different values of $\theta= \frac{\pi}{2}$, $\frac{\pi}{4}$, $\frac{\pi}{6}$, and $\frac{10\pi}{22}$, when $a=0.5$, and (b) at three different values of $a= 0.1$, $0.5$, and $0.998$, when $\theta=\frac{\pi}{4}$.}%
  \label{fig1}
\end{center}
\end{figure}

The four-vector gravitational potential $B^d$ can be calculated in the unit of gravitational radius $r_g$  as
\begin{equation}
B^d=\left[\epsilon^{abcd}e_{b\lambda}\left(\partial_c e_a^{\lambda} + \Gamma_{\alpha\mu}^{\lambda} e_c^{\alpha}e_a^{\mu}\right)\right](\hbar c/r_g),
\label{1.10}
\end{equation}
which leads to $B^0=B^3=0$ and $B^1\neq0$, $B^2\neq 0$. Here, $r_g=G M_B/c^2$ where we choose $M_B=10^{17}\times5.61\times10^{35} 
 \text{eV}$ is the mass of the PBH and $G=6.7\times 10^{-57}\rm\,{eV^{-2}}$.
Thus, the magnitude of gravitational vector potential can be written as
\begin{equation}
|\textbf{B}|=\sqrt{({B_1})^2+({B_2})^2}\rm\,{eV}.
\label{1.11}
\end{equation}
We find that when $a=0$, both $B_1$ and $B_2$ vanish. However, at $\theta=\pi/2$ and $a\neq0$, $B_1$ vanishes while $B_2$ is non-zero. For other values of $\theta$, both $B_1$ and $B_2$ are generally non-zero. In Fig.~\ref{fig1}, we illustrate the magnitude of the gravitational vector potential $|\textbf{B}|~\rm\,{(eV)}$ as a function of the radial distance $r$ from the black hole. As elucidated in Fig.~\ref{fig1}(a) and Fig.~\ref{fig1}(b), $|\textbf{B}|$ exhibits a monotonic decrease when radial distance increases from the black hole horizon. Particularly, Fig.~\ref{fig1}(a) shows $|\textbf{B}|$ vs $r$ for different values of $\theta$, with the specific angular momentum fixed at  $a=0.5$. We observe that $|\textbf{B}|$ across the range of $r$ is highest at $\theta=\pi/4$ (solid line) compared to its value at other $\theta$. Moreover, Fig.~\ref{fig1}(b) depicts $|\textbf{B}|$ vs $r$ for different values of specific angular momentum $a$, with $\theta$ fixed at $\pi/4$. As $a$ increases from a low (dashed line) to a high (solid line) value, $|\textbf{B}|$ remains high across the entire range of radial distance $r$ for a large value of $a$ (solid line). Hence, the strength of the four-vector gravitational potential strongly depends on the angle ($\theta$) with respect to the spin axis of the PBH and its specific angular momentum ($a$).

\section{Effective mass matrix of the neutrino-antineutrino system}
\label{Sect3}
By treating neutrinos as a left-handed particle, the spinor for a Majorana neutrino in the Weyl representation can be expressed as follows \cite{Sinha:2007uh}
\begin{equation}
    |\Psi\rangle=\begin{pmatrix}
        |\psi^c\rangle \\
        |\psi\rangle
    \end{pmatrix},
    \label{1.12}
\end{equation}
where $|\psi^c\rangle$ and $|\psi\rangle$ denote the antineutrino and neutrino spinors, having the lepton number eigenvalues -1 and 1, respectively. The Majorana mass ($m$) in terms of two-component spinors can be expressed as

\begin{equation}
    \bar{\Psi}\mathcal{M}\Psi=\begin{pmatrix}
        {\psi^c}^\dagger & \psi^\dagger
    \end{pmatrix}\begin{pmatrix}
        0 & -m\\
        -m & 0
    \end{pmatrix}\begin{pmatrix}
        \psi^c\\
        \psi
    \end{pmatrix}.
    \label{1.13}
\end{equation}
Now in gravitational field the Lagrangian density can be written as
\begin{align}
    (-g)^{-1/2}\mathcal{L}&=\begin{pmatrix}
        {\psi^c}^{\dagger} & {\psi}^{\dagger}
    \end{pmatrix}{i}\gamma^0\gamma^d\mathcal{\overleftrightarrow{\partial}}_d\begin{pmatrix}
        {\psi^c}\\ {\psi}
    \end{pmatrix}+\begin{pmatrix}
        {\psi^c}^{\dagger} & {\psi}^{\dagger}
\end{pmatrix}\gamma^0 B_1 \gamma^1 \gamma^5\begin{pmatrix}
        {\psi^c}\\ {\psi}
    \end{pmatrix}\nonumber\\
   & +\begin{pmatrix}
        {\psi^c}^{\dagger} & {\psi}^{\dagger}
\end{pmatrix}\gamma^0 B_2 \gamma^2 \gamma^5\begin{pmatrix}
        {\psi^c}\\ {\psi}
    \end{pmatrix}-\begin{pmatrix}
        {\psi^c}^{\dagger} & {\psi}^\dagger  \end{pmatrix}\gamma^0m\begin{pmatrix}
        {\psi^c}\\ {\psi}
    \end{pmatrix}.
    \label{1.14}
\end{align}
 In the Weyl or Chiral basis,  Eq.~(\ref{1.14}) can be re-expressed as
\begin{align}
    (-g)^{-1/2}\mathcal{L}=\begin{pmatrix}
        {\psi^c}^{\dagger} & {\psi}^{\dagger}
    \end{pmatrix}{i}\gamma^0\gamma^d\mathcal{\overleftrightarrow{\partial}}_d\begin{pmatrix}
        {\psi^c}\\ {\psi}
    \end{pmatrix}&\nonumber\\+\begin{pmatrix}
        {\psi^c}^{\dagger} & {\psi}^{\dagger}
\end{pmatrix}\begin{pmatrix}
    B_1\sigma^1+B_2\sigma^2 & 0\\
    0 & B_1\sigma^1+B_2\sigma^2
\end{pmatrix}\begin{pmatrix}
        {\psi^c}\\ {\psi}
    \end{pmatrix}&\nonumber\\
    +\begin{pmatrix}
        {\psi^c}^{\dagger} & {\psi}^\dagger  \end{pmatrix}\begin{pmatrix}
            0 & -m\\
            -m & 0
        \end{pmatrix}\begin{pmatrix}
        {\psi^c}\\ {\psi}
    \end{pmatrix}.
    \label{1.15}
\end{align}
The four terms in the above equation $B_1\sigma^1{\psi^c}^\dagger\psi^c$, $B_1\sigma^1{\psi}^\dagger\psi$, $B_2\sigma^2{\psi^c}^\dagger\psi^c$, and $B_2\sigma^2{\psi}^\dagger\psi$ do not violate lepton number and contribute effectively to the $\it{mass}$ of a Majorana neutrino. Therefore, writing the $\it{mass}$ terms together, the Euler-Lagrangian equation for neutrino and antineutrino in the background gravitational field is 

\begin{equation}
   i \gamma^0\gamma^d\mathcal{\overrightarrow{\partial}}_d\begin{pmatrix}
        {\psi^c}\\ {\psi} \end{pmatrix}+ \begin{pmatrix}
    B_1\sigma^1+B_2\sigma^2 & -m\\
    -m & B_1\sigma^1+B_2\sigma^2
\end{pmatrix}\begin{pmatrix}
        {\psi^c}\\ {\psi}
    \end{pmatrix}=0.
    \label{1.16}
\end{equation}
Hence, the effective mass matrix of the neutrino-antineutrino system is
\begin{equation}
    \mathcal{M}= \begin{pmatrix}
   B_1\sigma^1+B_2\sigma^2 & -m\\
    -m & B_1\sigma^1+B_2\sigma^2
\end{pmatrix}.
\label{1.16(a)}
\end{equation}

The eigenvalues of $\sigma^1$ and $\sigma^2$ are $\pm 1$. The effective mass matrix of the neutrino-antineutrino system in the presence of the gravitational field can be obtained as

\begin{equation}
    \mathcal{M}=\begin{pmatrix}
    -(B_1+B_2) & -m\\
    -m & (B_1+B_2)\\
  \end{pmatrix},
  \label{1.17}
\end{equation}
which has eigenvalues

\begin{equation}
    m_{1,2}=\mp\sqrt{(B_1+B_2)^2+m^2}.
    \label{1.17(a)}
\end{equation}

Thus, if one assumes that neutrinos solely possess Majorana-type masses, they acquire lepton number non-violating masses in addition that are equal in magnitude but opposite in sign while propagating in a gravitational field. Additionally, in this scenario, $\psi$ is not a mass eigenstate. 

We have assumed the effective mass matrix to have the form given in Eq.~(\ref{1.17}). In the future, we will show that the general effective mass matrix given in Eq.~(\ref{1.16(a)}) also leads to similar results.

\section{Neutrino-antineutrino oscillation in curved spacetime}\label{Sect4}
The effective mass matrix ($\mathcal{M}$), given by Eq.~(\ref{1.17}), is Hermitian and can be diagonalize by a unitary transformation $\mathbf{U}(\theta_\nu)$. Motivated by the nature of neutral kaon system, we define two mass eigenstates such that
\begin{equation}
\begin{pmatrix}
    |{\nu_1}\rangle\\
    |{\nu_2}\rangle
\end{pmatrix}=\mathbf{U}(\theta_\nu)\begin{pmatrix}
    |\psi^c\rangle\\
    |{\psi}\rangle
\end{pmatrix},\hspace{0.3cm} \mathbf{U}(\theta_\nu)=\begin{pmatrix}
    \cos\theta_\nu & \sin\theta_\nu\\
    -\sin\theta_\nu & \cos\theta_\nu
\end{pmatrix},
\label{1.19}
    \end{equation}
with mixing angle
\begin{equation}
    \theta_\nu=\tan^{-1}\left[\frac{m}{(B_1+B_2)+\sqrt{(B_1+B_2)^2+m^2}}\right].
    \label{1.20}
\end{equation}
When $m=0$, no mixing possible. $(B_1+B_2)\rightarrow0$ implies $\theta_\nu\rightarrow\frac{\pi}{4}$. We show in Fig.~\ref{fig2}(a) how the mixing angle $\theta_\nu$ changes with radial distance $r$ from the black hole for a low (dashed line) and a high (solid line) value of specific angular momentum $a$, keeping $\theta$ fixed at $\frac{\pi}{4}$ and using the Majorana mass\footnote{The Majorana mass used here is calculated in the next section.} $m=0.0497666\rm\,{eV}$. We observe that the behavior of $\theta_\nu$ is sharp for a small value (dashed line) of $a$ and becomes shallow for a large value (solid line). At a small radial distance $r$ from the black hole, the magnitude of gravitational vector potential is very high (see Fig.~\ref{fig1}(b)), causing $\theta_\nu$ to approach a low value for both values of $a$. Conversely, at a large radial distance $r$ from the black hole, where the magnitude of gravitational vector potential is low, $\theta_\nu$ saturates to $\pi/4$ for both values of $a$. 
\def\figsubcap#1{\par\noindent\centering\footnotesize(#1)}
\begin{figure}[t]%
\begin{center}
  \parbox{2.1in}{\includegraphics[width=2.2in]{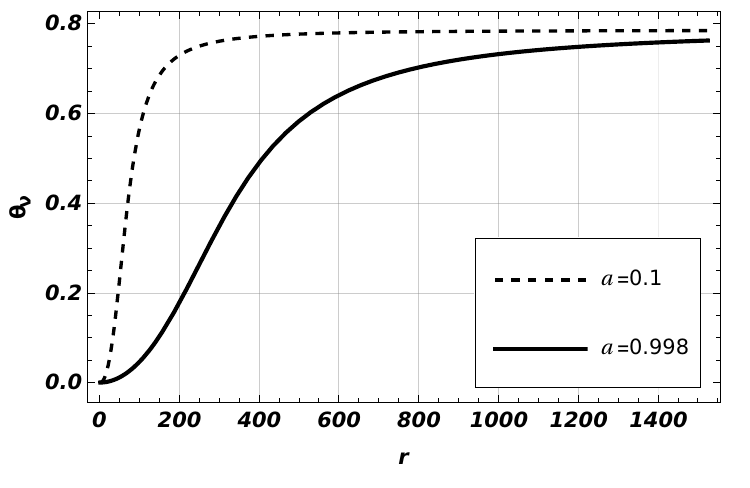}
  \figsubcap{a}}
  \hspace*{4pt}
  \parbox{2.1in}{\includegraphics[width=2.2in]{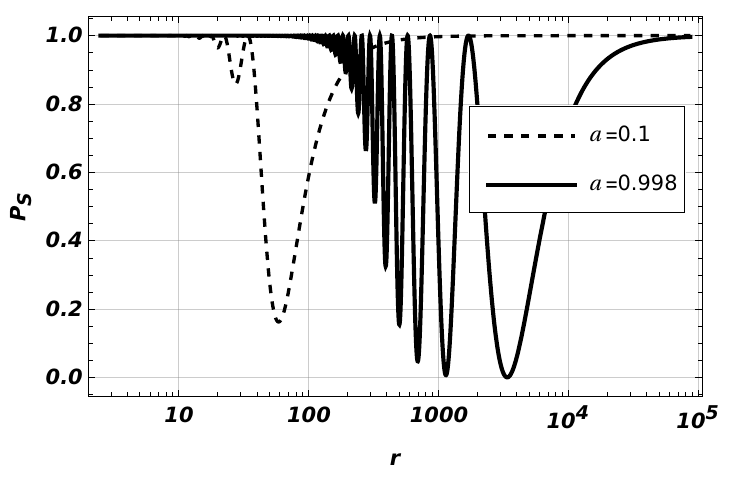}
  \figsubcap{b}}
  \caption{(a) Mixing angle $\theta_\nu$ and (b) the survival probability $P_s$ of the initial state $|{\nu_1}\rangle$ are shown as a function of radial distance $r$ from the black hole at a low (dashed line) and a high value (solid line) of specific angular momentum ($a$), when $\theta=\frac{\pi}{4}$ and $m=0.0497666\rm\,{eV}$.}%
  \label{fig2}
\end{center}
\end{figure}

The time evolution of the particle and antiparticle is
\begin{equation}
|{\psi(t)}\rangle\rightarrow|{\psi(0)}\rangle e^{-iE_{\nu} t},\hspace{0.5cm}|{\psi^c(t)}\rangle\rightarrow|{\psi^c(0)}\rangle e^{-iE_{\nu^c}t}.
\label{1.20}
\end{equation}
 Due to the gravitational interaction, neutrino-antineutrino states couple together with modified energy 
  \begin{equation}
E_\nu=\sqrt{(\vec{p}-\vec{B})^2+m^2}\,,\hspace{0.5cm}
E_{\nu^c}=\sqrt{(\vec{p}+\vec{B})^2+m^2}\,,
\label{1.21}
  \end{equation}
  where $\vec{p}$ is momentum\footnote{In the relativistic limit $|\vec{p}|>>m$, we assume $|\vec{p}|\approx 1\rm eV$.}. As a consequence of the difference in energy (effective mass), the time evolution of the neutrino and antineutrino differs, leading to possible oscillations in the neutrino-antineutrino system. The survival probability of $|{\nu_1}\rangle$ at any time $t$ can be expressed as
\begin{equation}
    P_s(t)=1-\sin^2(2\theta_\nu) \sin^2\left[\frac{1}{2}(E_\nu-E_{\nu^c})t\right].
   \label{1.22}
\end{equation}
The oscillation probability of the initial state $|\nu_1\rangle$ is $P_o(t)=1-P_s(t)$. Thus, when $P_s(t)\neq 0$ implies $P_o(t)\neq 0$. However, in the relativistic limit, for an observer at infinity, it is essential to relate the neutrino propagation time $t$ to the radial distance $r$ from the black hole. To achieve this, we consider neutrinos traveling along radial trajectories. We derive the relationship between $t$ and $r$ from the line element given by Eq.~(\ref{1.7}) by setting $ds^2=0$. Additionally, we set $d\phi=0$ and $d\theta=0$. Consequently, the relationship we obtain between  $t$ and $r$ is as follows
\begin{equation}
    t=\int^r_{r_1> r_{es}}\left[\frac{(r^2+a^2\cos^2\theta)}{\sqrt{(r^2-2r+a^2)(r^2-2r+a^2\cos^2\theta)}}\right]dr,
    \label{1.23}
\end{equation}
where $r_{\rm es}$ is the radius of the Ergosphere of the spinning black hole which can be denoted as $r_{es}=1+\sqrt{(1-a^2\cos^2\theta)}$. Subsequently, using Eq.~(\ref{1.23}) in Eq.~(\ref{1.22}), the survival probability of the initial state $|{\nu_1}\rangle$ at any time $t$ can be expressed as a function of radial distance $r$ from the black hole as
    \begin{equation}
P_s(r)=1-\sin^2(2\theta_\nu)\sin^2\left[\left(\frac{E_\nu-E_{\nu^c}}{2}\right)\int^r_{r_1> r_{es}}f(r,\theta,a)dr\right],
\label{1.24}
    \end{equation}
where 
\begin{equation*}
f(r,\theta,a)=\frac{(r^2+a^2\cos^2\theta)}{\sqrt{(r^2-2r+a^2)(r^2-2r+a^2\cos^2\theta)}}.
\end{equation*}
Fig.~\ref{fig2}(b) shows the survival probability $P_s$ of the initial state $|{\nu_1}\rangle$  as a function of radial distance $r$ for a low (dashed line) and a high (solid line) value of the specific angular momentum $a$, with $\theta=\frac{\pi}{4}$ and $m=0.0497666\rm\,{eV}$. We observe that the peak of $P_s$ varies when the strength of $a$ changes from a low (dashed line) to a high value (solid line) near the black hole. At low $a$ (dashed line), $\theta_\nu$ saturates quickly to $\frac{\pi}{4}$, and the energy difference in the oscillatory term (see Eq.~({\ref{1.24}})) is also very less, so $P_s$ saturates to 1 quickly. However, at large $a$ (solid line), $\theta_\nu$ saturate to $\frac{\pi}{4}$ slowly, and the energy difference in the oscillatory term is sustained for a large range of $r$; therefore, the peak in $P_s$ changes between 1 and 0 frequently near the black hole. This suggests that the primary cause of neutrino-antineutrino transitions near the black hole is the change in the strength of the gravitational vector potential. However, as we go very far away from the black hole, at the large $a$ (solid line), the $\theta_\nu$ saturate to $\frac{\pi}{4}$, and the energy difference becomes minimal, causing the oscillatory terms in Eq.~({\ref{1.24}}) to vanish, i.e., $\sin^2\left[\left(\frac{E_\nu-E_{\nu^c}}{2}\right)\int^r_{r_1> r_{es}}f(r,\theta, a)dr\right]\rightarrow 0$. Consequently, $P_s(r)$ in Eq.~(\ref{1.24}) tends to 1. These conditions make neutrinos less prone to oscillation over very large radial distances $r$ from the black hole. 

The neutrino-antineutrino oscillations in curved spacetime presented here serve as a simplified model. A more comprehensive analysis would take into account two-flavor oscillations of the neutrino-antineutrino system in curved spacetime. 

\section{Two-flavor oscillation with neutrino-antineutrino mixing in curved spacetime}\label{Sect5}
In Ref.\cite{Sinha:2007uh} the construction of the Majorana mass ($m$) in terms of four-component spinors is given in the two-flavor scenarios of the neutrino-antineutrino system.
Following the method provided in Sec.~\ref{Sect3}, in the background gravitational field, the Euler-Lagrangian equation in four spinor components forms a Majorana neutrino as
\begin{equation}
    i\gamma^0\gamma^d \overrightarrow{\partial}_d\begin{pmatrix}
        \psi^c_e\\
        \psi^c_\mu\\
        \psi_e\\
        \psi_\mu\\
    \end{pmatrix}+\mathcal{M}_4\begin{pmatrix}
        \psi^c_e\\
        \psi^c_\mu\\
        \psi_e\\
        \psi_\mu\\
    \end{pmatrix}=0,
    \label{1.25}
    \end{equation}
where $\psi_e,\psi_\mu,\psi^c_e,\psi^c_\mu$ are corresponding flavor spinors for electron neutrino, muon neutrino, electron antineutrino, and muon antineutrino, respectively. Here, $\mathcal{M}_4$ in the case of the two-flavor scenario is the most general flavor mixing mass matrix in the presence of gravity and can be expressed as
\begin{equation}
    \mathcal{M}_4=\begin{pmatrix}
        -(B_1+B_2)\mathbf{I} & -\mathbf{M}\\
        -\mathbf{M} & (B_1+B_2)\mathbf{I}\\
    \end{pmatrix},
    \label{1.26}
\end{equation}
where $\mathbf{I}$ is the $2\times2$ unit matrix and 
\begin{equation}
    \mathbf{M}=\begin{pmatrix}
        m_e & m_{e\mu}\\
        m_{e\mu} & m_\mu\\
    \end{pmatrix}\equiv \mathbf{U(\theta_{vac})} . \rm {diag(m_1,m_2)}. \mathbf{U^\dagger(\theta_{vac})}.
    \label{1.27}
\end{equation}
Here, the Majorana masses for the electron and muon neutrino are $m_{e}$ and $m_{\mu}$, respectively. $m_{e\mu}$ is the Majorana mixing mass. The neutrino masses and the mixing matrix in vacuum are denoted by $m_{1,2}$ and $\mathbf{U(\theta_{vac})}$ (where $\theta_{\text{vac}}$ is the mixing angle in vacuum), respectively. However, for the sake of simplification of our analysis, we assume that the $4\times4$ mass matrix can be represented in a $2\times2$ block form to determine the effective masses of the mass eigenstates. We use the method outlined in Sec.~IV of Ref. \cite{Sinha:2007uh}.

The effective mass matrix $\mathcal{M}_4$ is Hermitian and can be diagonalise by the unitary matrix\cite{Sinha:2007uh} $\mathbf{T}$, where
\begin{equation}
\mathbf{T}=\begin{pmatrix}
    \cos\theta_e \cos\phi_1 & -\cos\theta_e \sin\phi_1 & -\sin\theta_e \cos\phi_2 & \sin\theta_e \sin\phi_2\\
        \cos\theta_\mu \sin\phi_1 & \cos\theta_\mu \cos\phi_1 & -\sin\theta_\mu \sin\phi_2 & -\sin\theta_\mu \cos\phi_2\\
       \sin\theta_e \cos\phi_1 & -\sin\theta_e \sin\phi_1 & \cos\theta_e \cos\phi_2 & -\cos\theta_e \sin\phi_2\\
        \sin\theta_\mu \sin\phi_1 & \sin\theta_\mu \cos\phi_1 & \cos\theta_\mu \sin\phi_2 & \cos\theta_\mu \cos\phi_2\\
   \end{pmatrix}.
  \label{1.28}
\end{equation}
Thus, the flavor and mass eigenstates ($|\chi_i\rangle$) are related via $\mathbf{T}$ as \cite{Sinha:2007uh}
\begin{equation}
    \begin{pmatrix}
       |{\psi_e^c}\rangle\\
        |{\psi_\mu^c}\rangle\\
        |{\psi_e}\rangle\\
        |{\psi_\mu}\rangle\\
\end{pmatrix}=\mathbf{T}\begin{pmatrix}
        |{\chi_1}\rangle\\
        |{\chi_2}\rangle\\
        |{\chi_3}\rangle\\
        |{\chi_4}\rangle\\
\end{pmatrix}.
   \label{1.29}
\end{equation}
The energy corresponding to mass eigenstates with momentum $\vec{p}$ are 
\begin{align}
E_1&=\sqrt{(\vec{p}+\vec{B})^2+M_1^2},\hspace{0.3cm}E_2=\sqrt{(\vec{p}+\vec{B})^2+M_2^2},\nonumber\\
   E_3&=\sqrt{(\vec{p}-\vec{B})^2+M_3^2},\hspace{0.3cm}
   E_4=\sqrt{(\vec{p}-\vec{B})^2+M_4^2}.
   \label{1.30}
\end{align}
The masses corresponding to the mass eigenstates are given as
\begin{eqnarray}
    {M_{1,2}=\frac{1}{2}[(m_{e1}+m_{\mu 1})\pm\sqrt{(m_{e1}-m_{\mu 1})^2+4m^2_{e\mu}}],}&\nonumber\\
    {M_{3,4}=\frac{1}{2}[(m_{e2}+m_{\mu 2})\pm\sqrt{(m_{e2}-m_{\mu 2})^2+4m^2_{e\mu}}],}& 
   \label{1.31}
\end{eqnarray}
with
\begin{equation}
    m_{(e,\mu)1}=-\sqrt{(B_1+B_2)^2+m^2_{e,\mu}}\,, \hspace{0.5cm} m_{(e,\mu)2} =\sqrt{(B_1+B_2)^2+m^2_{e,\mu}}\,,
    \label{1.32}
\end{equation}
where $m_{(e,\mu)1}$ and $m_{(e,\mu)2}$ are expressed using Eq.~(\ref{1.17(a)}).
\def\figsubcap#1{\par\noindent\centering\footnotesize(#1)}
\begin{figure}[t]%
\begin{center}
  \parbox{2.1in}{\includegraphics[width=2.2in]{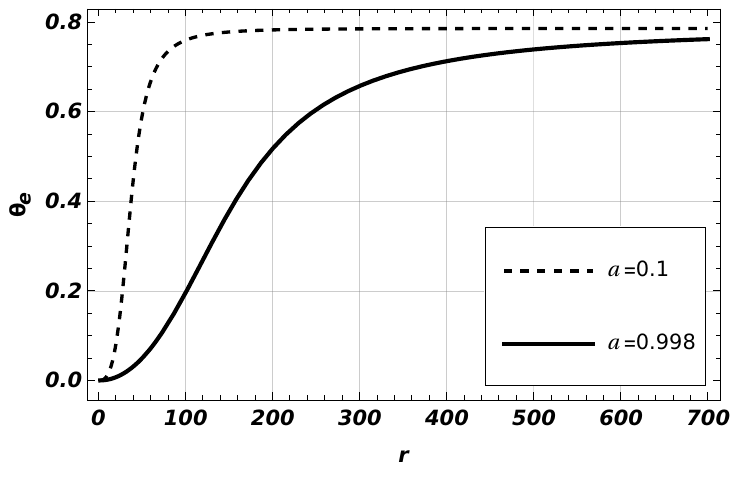}
  \figsubcap{a}}
  \hspace*{4pt}
  \parbox{2.1in}{\includegraphics[width=2.2in]{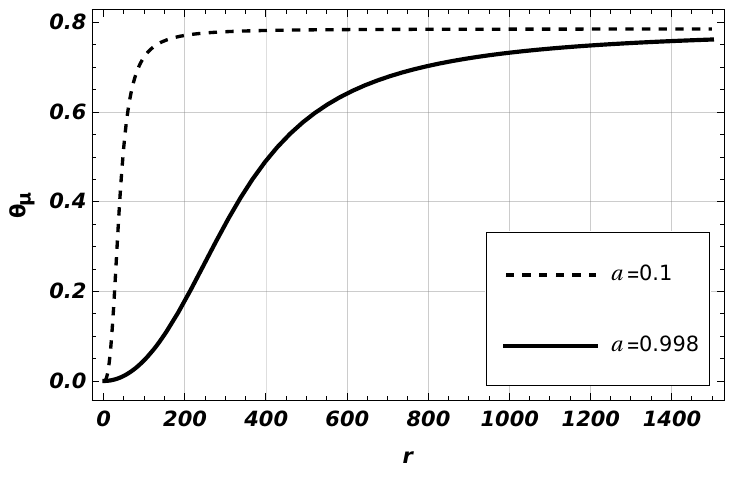}
  \figsubcap{b}}
  \caption{In the two-flavor scenario, mixing angles (a) $\theta_e$ and (b) $\theta_\mu$ are shown as a function of radial distance $r$ from the black hole at a low (dashed line) and a high value (solid line) of specific angular momentum ($a$), when $\theta=\frac{\pi}{4}$, $m_e=0.0497666\rm\,{eV}$, and $m_\mu=0.0500574\rm\,{eV}$.}%
  \label{fig3}
\end{center}
\end{figure}

\def\figsubcap#1{\par\noindent\centering\footnotesize(#1)}
\begin{figure}[t]%
\begin{center}
  \parbox{2.1in}{\includegraphics[width=2.2in]{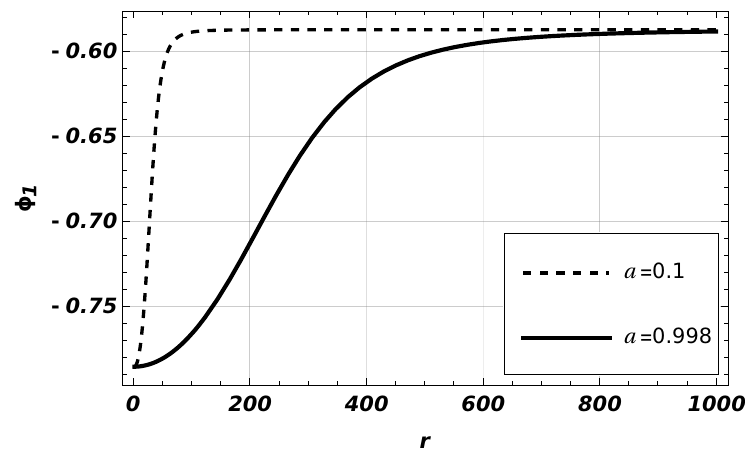}
  \figsubcap{a}}
  \hspace*{4pt}
  \parbox{2.1in}{\includegraphics[width=2.2in]{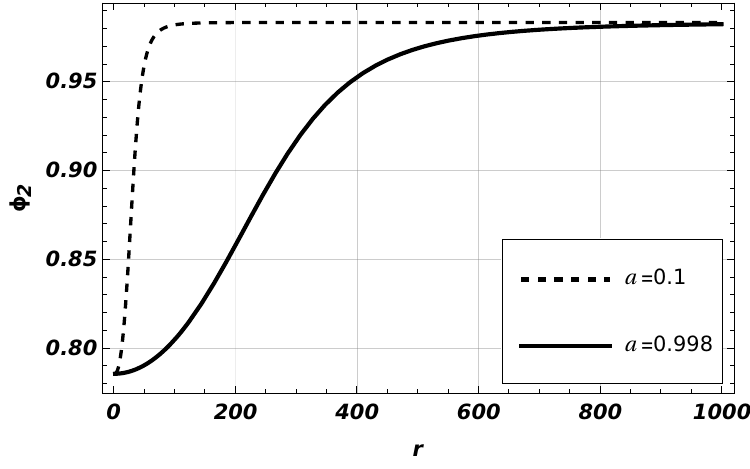}
  \figsubcap{b}}
  \caption{In the two-flavor scenario, mixing angles (a) $\phi_1$ and (b) $\phi_2$ are shown as a function of radial distance $r$ from the black hole at a low (dashed line) and a high value (solid line) of specific angular momentum ($a$), when $\theta=\frac{\pi}{4}$, $m_e=0.0497666\rm\,{eV}$, $m_\mu=0.0500574\rm\,{eV}$, and $m_{e\mu}=0.000347999\rm\,{eV}$.}%
  \label{fig4}
\end{center}
\end{figure}
The mixing angles for neutrino-antineutrino and electron-muon neutrino mixing are related to the masses and the gravitational potential, respectively, given by
\begin{equation}
    \theta_{e,\mu}=\tan^{-1}\left[\frac{m_{e,\mu}}{(B_1+B_2)+\sqrt{(B_1+B_2)^2+m^2_{e,\mu}}}\right],
    \label{1.33}
\end{equation}
\begin{equation}
    \phi_{1,2}=\tan^{-1}\left[\frac{\mp 2 m_{e\mu}}{m_{e(1,2)}-m_{\mu(1,2)}+\sqrt{(m_{e(1,2)}-m_{\mu(1,2)})^2+4m^2_{e\mu}}}\right].
    \label{1.34}
\end{equation}
\begin{table}       %Table~1
\tbl{The values of the neutrino mixing parameters for Inverted Hierarchy (IH) that we considered in our analysis are taken from the PDG Ref.\cite{Navas:2024} along with their corresponding $1\sigma$ errors ($90\%$ CL).}
{\begin{tabular}{@{}cc@{}}
\hline\\
&\\[-15pt]
Parameters & Best fit$\pm 1\sigma$ \\[3pt]
\hline\\
&\\[-15pt]
$\Delta m^2_{21}$ & $(7.53\pm0.18)\times 10^{-5}\rm\,{eV^2}$ \\[8pt]
$\Delta m^2_{32}$ & $(-2.529 \pm 0.029)\times 10^{-3}\rm\,{eV^2}$ \\[8pt]
$\theta_{\text{vac}}$  & $(33.66\pm 0.72)^\circ$  \\[8pt]
\hline
\end{tabular} \label{tp1}}
\end{table}
Furthermore, the spectrum of neutrino masses can be classified into two equivalent orderings. The first is known as Normal Ordering (NO), where $m_1<m_2<m_3$, while the second is Inverted Ordering (IO), with $m_3<m_1<m_2$. The Particle Data Group (PDG) emphasizes the hierarchy present among the mass splittings \cite{Navas:2024}, $\Delta m^2_{21}<<|\Delta m^2_{31}|\simeq|\Delta m^2_{32}|$ with $\Delta m^2_{ij}\equiv m^2_i-m^2_j$. 
It is important to note that $\Delta m^2_{32}>0$ and $\Delta m^2_{32}<0$ corresponds to NO and IO, respectively. Depending on the value of the lightest neutrino mass, the neutrino mass spectrum can be further categorized into a Normal Hierarchical (NH) spectrum ($m_1<<m_2<m_3$) and Inverted Hierarchical (IH) spectrum ($m_3<<m_1<m_2$). In the IH spectrum, the neutrino mixing parameters $\Delta m^2_{ij}$ and $\theta_{\text{vac}}$, along with
their $1\sigma$ best-fit data, is 
presented in Table~\ref{tp1}. Subsequently, we get $m_1\simeq\sqrt{|\Delta m^2_{32}+\Delta m^2_{21}|}\sim 0.0495\pm 3.1 \times 10^{-4}\rm\, {eV}$ and $m_2\simeq\sqrt{|\Delta m^2_{32}|}\sim 0.05\pm 2.867\times 10^{-4}\rm\,{eV}$. Using the $m_1$, $m_2$, and the vacuum mixing angle $\theta_{\text{vac}}$ in Eq.~(\ref{1.27}), we calculate the value of Majorana mass\footnote{In this paper, we use the calculated values of the Majorana mass.} as $m_e=0.0497666\pm 3.115\times 10^{-4}\rm\,{eV}$, $m_\mu=0.0500574\pm 3.02\times 10^{-4}\rm\,{eV}$, and $m_{e\mu}=0.000347999\pm 1.01\times 10^{-5}\rm\,{eV}$. Thus, using Eq.~(\ref{1.33}) and Eq.~(\ref{1.34}), in Figs.~\ref{fig3}(a), $\ref{fig3}(b)$, $\ref{fig4}(a)$, and $\ref{fig4}(b)$, we illustrate the behavior of the mixing angle $\theta_e$, $\theta_\mu$, $\phi_1$, and $\phi_2$, respectively, as functions of radial distance $r$ for a low (dashed line) and a high (solid line) value of specific angular momentum $a$. 

Moreover, the time evolution of the electron flavor neutrino state in a linear superposition of flavor basis can be expressed as
\begin{equation}
   |{\psi_e(t)}\rangle=T_{e e^c}(t)|{\psi^c_e}\rangle +T_{e \mu^c}(t)|{\psi^c_\mu}\rangle
   + T_{ee}(t) |{\psi_e}\rangle+T_{e \mu}(t) |{\psi_\mu}\rangle.
   \label{1.35}
\end{equation}
\def\figsubcap#1{\par\noindent\centering\footnotesize(#1)}
\begin{figure}[t]%
\begin{center}
  \parbox{2.1in}{\includegraphics[width=2.2in]{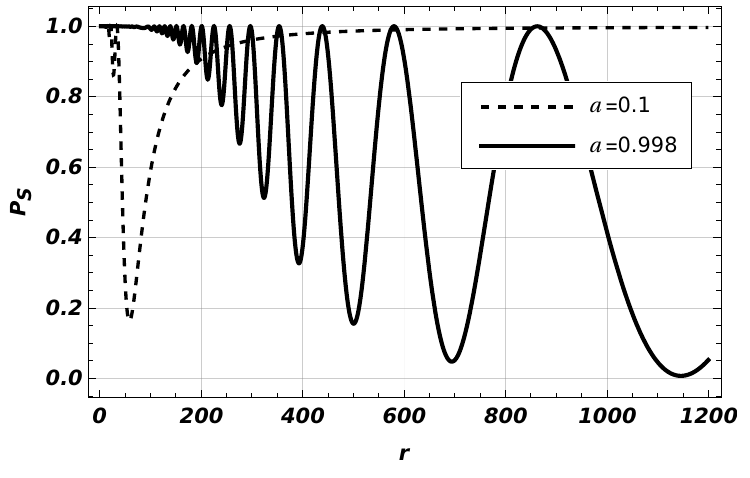}
  \figsubcap{a}}
  \hspace*{4pt}
  \parbox{2.1in}{\includegraphics[width=2.2in]{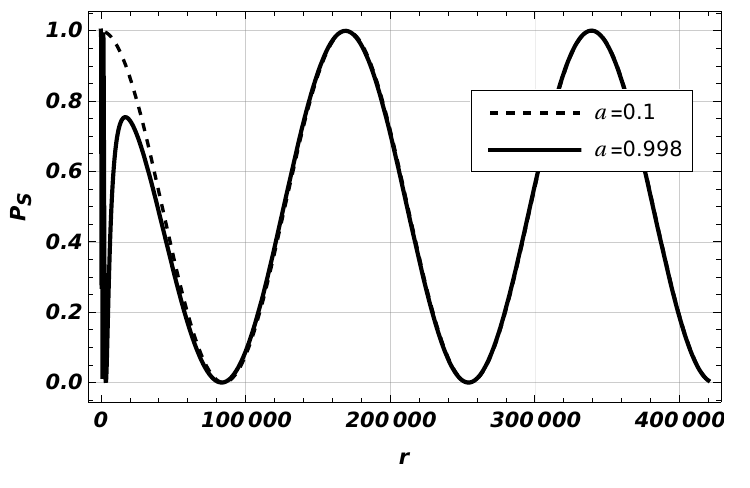}
  \figsubcap{b}}
  \caption{In the two-flavor scenario, the survival probability $P_s$ of the initial state $|\psi_e\rangle$ is shown as a function of (a) small range and (b) large range of radial distance $r$ from the black hole, at a low (dashed line) and a high value (solid line) of specific angular momentum ($a$), when $\theta=\frac{\pi}{4}$, $m_e=0.0497666\rm\,{eV}$, $m_\mu=0.0500574\rm\,{eV}$, and $m_{e\mu}=0.000347999\rm\,{eV}$.}%
  \label{fig5}
\end{center}
\end{figure}
The survival probability of the initial state $|{\psi_e}\rangle$ is
    \begin{align} 
    P_s(t)&=|T_{ee}(t)|^2=|\cos^2\phi_2 (e^{-i E_3 t} \cos^2\theta_e + e^{-i E_1 t} \sin^2\theta_e)\nonumber\\& + \sin^2\phi_2 (e^{-i E_4 t} \cos^2\theta_\mu + e^{-i E_2 t} \sin^2\theta_\mu)|^2 .
    \label{1.36}
    \end{align}

In the relativistic limit, using Eq.~(\ref{1.23}) in Eq.~(\ref{1.36}), Fig.~{\ref{fig5}(a)} shows the survival probability $P_s$ as a function of the small range of radial distance $r$ from the black hole for a low (dashed line) and a high (solid line) value of specific angular momentum $a$. We observe that near the black hole, the peak of survival probability $P_s$ fluctuates more frequently between 1 and 0 for a high $a$ (solid line) compared to a low $a$ (dashed line). This occurs because the strength of the gravitational vector potential is greater for a high value of $a$ (solid line) at a given radial distance $r$ as previously depicted in Fig.~\ref{fig1}(b). However, Fig.~{\ref{fig5}(b)} illustrates that as we move far away from the black hole, the strength of the gravitational vector potential decreases, leading to a negligible effect of the specific angular momentum $a$ on the survival probability $P_s$. Consequently, at large radial distances $r$ from the black hole, the two-flavor oscillations of the neutrino-antineutrino system begin to propagate in a vacuum.

\section{Quantum speed limit in flavor oscillations of neutrino-antineutrino system in curved spacetime}\label{Sect6}
The quantum speed limit (QSL) concept originated from the uncertainty relationship between conjugate variables in quantum mechanics  \cite{Mandelstam:1991,Margolus:1997ih}. It represents a fundamental constraint set by quantum mechanics on the rate of evolution for any quantum system undergoing a specific dynamical process. Thus, QSL establishes a minimum time necessary for a quantum system to transition from its initial state to its final state \cite{Thakuria:2022taf}. Recently, Ref.~\cite{Bouri:2024kcl} conducted a thorough analysis of the QSL time in the framework of neutrino oscillations in matter. 

The formulation of the QSL time ($T_{QSL}$) is influenced by factors such as the Bures angle ($S_0$), representing the shortest (or geodesic) distance between the initial and final states, and the variance (or fluctuations) in the driving Hamiltonian ($\Delta H$). If the dynamic of the system is unitary and it is driven by the time-independent Hermitian Hamiltonian $H$, then the QSL time $T_{QSL}$ for such a system is defined by \cite{Thakuria:2022taf}
\begin{eqnarray}
T\geq T_{QSL};\hspace{0.5cm} T_{QSL}= \frac{\hbar S_0} {\Delta H},
\label{1.37}
\end{eqnarray}
where
\begin{equation}
    {S_0} = \cos^{-1}(|\langle \psi (0)|\psi(T)\rangle|),
    \label{1.38}
    \end{equation}
and
\begin{align}
\Delta H =\sqrt{\langle \psi|H^2|\psi\rangle - (\langle\psi|H|\psi\rangle )^2}
 =\sqrt{(\langle \dot{\psi} (t)|\dot{\psi}(t)\rangle-(i\langle \psi (t)|\dot{\psi}(t)\rangle)^2)}.
\label{1.39}
\end{align}
Here, $\dot{|{\psi(t)}\rangle}\equiv\frac{d|{\psi(t)}\rangle}{dt}$.
Eq.~(\ref{1.5}) shows that the Hamiltonian obtained from the Lagrangian equation is Hermitian and time-independent. Furthermore, the two-flavor oscillations of the neutrino-antineutrino system in curved spacetime obey unitary dynamics. Therefore, one can impose the QSL time technique during the flavor oscillations of the neutrino-antineutrino in the background of the gravitational field. 

Using Eq.~(\ref{1.23}) in Eq.~(\ref{1.36}), and substituting it in Eq.~(\ref{1.38}), we find the expression of the Bures angle ($S_0$) for the time evolved state $|\psi_e(t)\rangle$ (see Eq.~(\ref{1.35})), as a function of radial distance $r$  from the black hole as
    \begin{equation}
    {S_0}(r) = \cos^{-1}(\sqrt{P_s(r)})=\cos^{-1}(\sqrt{|T_{ee}(r)|^2}),
    \label{1.40}
    \end{equation}
   where $P_s=|\langle \psi_e (0)|\psi_e(r)\rangle|^2$ is the survival probability of the initial state $|\psi_e\rangle$.
   \def\figsubcap#1{\par\noindent\centering\footnotesize(#1)}
\begin{figure}[t]%
\begin{center}
  \parbox{2.1in}{\includegraphics[width=2.2in]{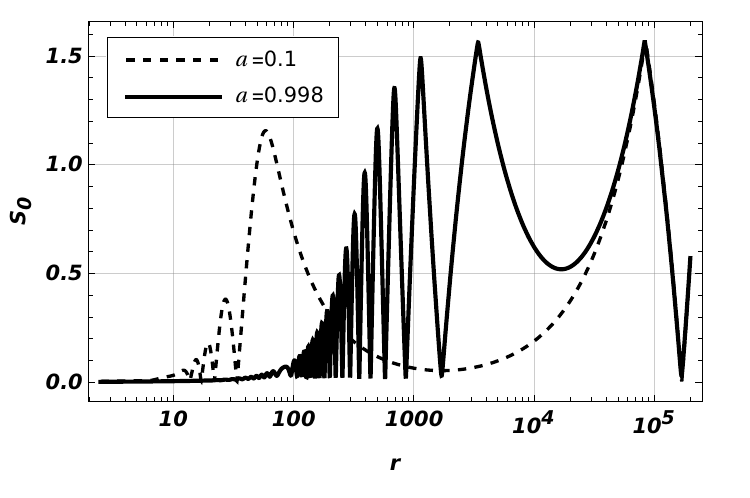}
  \figsubcap{a}}
  \hspace*{4pt}
  \parbox{2.1in}{\includegraphics[width=2.2in]{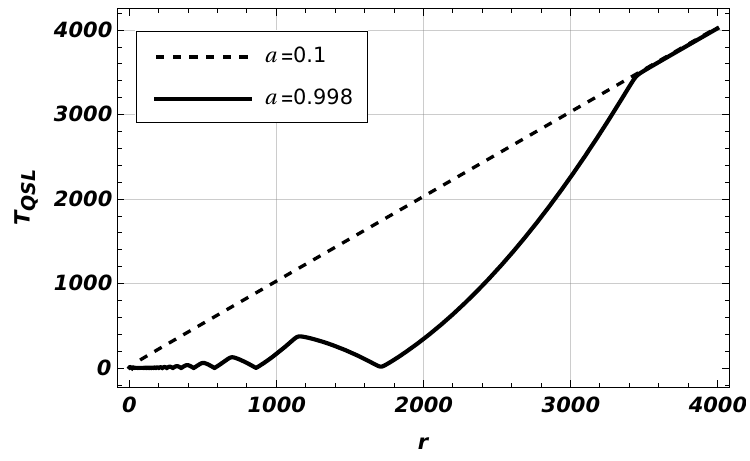}
  \figsubcap{b}}
  \caption{In the two-flavor scenario, (a) the Bures angle $S_0$ and (b) the QSL time $T_{QSL}$ of the initial state $|\psi_e\rangle$ are shown as a function of radial distance $r$ from the black hole at a low (dashed line) and a high value (solid line) of specific angular momentum ($a$), when $\theta=\frac{\pi}{4}$, $m_e=0.0497666\rm\,{eV}$, $m_\mu=0.0500574\rm\,{eV}$, and $m_{e\mu}=0.000347999\rm\,{eV}$.}%
  \label{fig6}
\end{center}
\end{figure}
   In Fig.~\ref{fig6}(a), we analyze the Bures angle $S_0$ as functions of radial distance $r$ at $\theta=\pi/4$, for a low (dashed line) and a high (solid line) value of the specific angular momentum $a$. Over a small range of radial distance $r$ from the black hole, the survival probability $P_s$ of the initial state $|\psi_e\rangle$
  varies significantly between 1 and 0 based on the strength of 
$a$ (see Fig.~{\ref{fig5}(a)}), which in turn leads to the expected behavior of  $S_0$
  as illustrated in Fig.~\ref{fig6}(a). However, at a very large radial distance 
$r$ from the black hole, the influence of gravity on $P_s$ diminishes, as shown in Fig.~{\ref{fig5}(b)}. This results in $S_0$
  in Fig.~\ref{fig6}(a) achieving its maximum value ($S_0\approx \frac{\pi}{2}$)
 and oscillating similarly for both low (dashed line) and high (solid line) values of $a$. 

 Furthermore, using Eq.~(\ref{1.35}) in Eq.~(\ref{1.39}), we compute energy fluctuation $\Delta H$ as a function of radial distance $r$. Subsequently, using Eq.~(\ref{1.40}) and Eq.~(\ref{1.37}), we estimate $T_{\text{QSL}}$ for the initial electron flavor neutrino state $|{\psi_{e}\rangle}$ as ($\hbar\approx1$)
\begin{equation}
    T_{QSL}(r)=\frac{\cos^{-1}(\sqrt{|T_{ee}(r)|^2})}{\Delta{H}(r)}.
    \label{1.41}
\end{equation}
Fig.~{\ref{fig6}(b)} depicts the QSL time $T_{QSL}$ of the initial state $|\psi_e\rangle$ as a function of radial distance $r$
for a low (dashed line) and a high (solid line) value of the specific angular momentum $a$. At a lower value of $a$ (dashed line), $T_{QSL}$ behaves tightly, according to the time bound condition, $\frac{T_{QSL}}{r}=1$. This implies that the evolution speed of the initial state $|\psi_e\rangle$ remains unchanged, maximizing the minimum time required for oscillation. Conversely, as $a$ increases (solid line), $T_{QSL}$ 
  decreases, indicating that the initial state $|\psi_e\rangle$
oscillates more rapidly. This can be reflected in the time-bound condition, $\frac{T_{QSL}}{r}<1$, which denotes a faster dynamical evolution of the quantum state. This behavior is expected since a larger specific angular momentum (solid line) is associated with a greater magnitude of the gravitational vector potential (see solid line in Fig.~{\ref{fig1}(b)}). Therefore, the QSL time for the two-flavor oscillation of the neutrino-antineutrino system in curved spacetime is significantly reduced with increasing specific angular momentum $a$. However, it's important to recognize that these findings are meaningful only within a limited range of radial distances from the black hole. Near the black hole horizon, the value of Bures angle ($S_0$) is very small for a high value of $a$ as illustrated in Fig.~\ref{fig6}(a). This suggests that the distance between initial and final states tends to zero i.e. the system maintains its initial configuration in the strong gravitational regime. Hence from the $T_{QSL}$ graph one can see that the time required to oscillate is almost zero for high $a$, near the black hole horizon. At a very large radial distance, the effects of gravity diminish, leading to time-bound condition $\frac{T_{QSL}}{r}\rightarrow 1$ independent of $a$, indicating a slower dynamical evolution of the initial state $|\psi_e\rangle$ as it approaches the maximum QSL time.

\section{Conclusions}\label{Sect7}
In this work, we have analyzed the dynamics of the neutrino-antineutrino system in the spacetime of a spinning primordial black hole (PBH) by considering the Dirac-Lagrangian formalism in curved spacetime. Boyer-Lindquist coordinates are used for a spinning black hole to obtain an analytical expression for the four-vector gravitational potential in the Hermitian Dirac Hamiltonian. We have observed that the magnitude of the gravitational vector potential highly depends on the angle the spinor makes with respect to the spin axis of the PBH and on its specific angular momentum. At a fixed angle (say $\theta=\pi/4$), a high value of specific angular momentum corresponds to a high magnitude of the gravitational vector potential. 

We have further shown that this gravitational vector potential contributes to the effective mass matrix of the neutrino-antineutrino system. We have studied the oscillation of the neutrino-antineutrino system in curved spacetime. As a specific example, we have considered a model where the neutrinos are radially moving away from the spinning PBH. As the magnitude of the gravitational vector potential is affected by a low to high value of the specific angular momentum of the black hole, we have observed that near the black hole, the transition probability between neutrino and antineutrino are also highly influenced by a low and a high value of the specific angular momentum. We further delve into the analysis for two-flavor oscillations of the neutrino-antineutrino system in curved spacetime. Near the black hole, depending on the strength of its specific angular momentum, the transition probability of the initial neutrino flavor state changes. However, as the spinor goes very far away from the black hole, the effect of the specific angular momentum (or the magnitude of the four-vector gravitational potential) becomes negligible on the neutrino transition probability. This causes the two-flavor oscillation of the neutrino-antineutrino system in the vacuum.

Furthermore, we have studied the effects of specific angular momentum of the PBH on the quantum speed limit (QSL) in two-flavor oscillations of the neutrino-antineutrino system. According to our finding, the QSL time has shown a rapid evolution for the initial neutrino flavor state at a high value of specific angular momentum near the black hole.

\section*{Acknowledgments}
The authors would like to acknowledge the project funded by SERB, India, with Ref. No. CRG/2022/003460, for supporting this research. AKJ expresses gratitude to ANRF (SERB), India, for providing financial assistance through the ITS scheme, which facilitated participation and presentation of this work at the Seventeenth Marcel Grossmann Meeting, held in Italy from July 7 to July 12, 2024.
%\appendix{Appendix}

\vfill
\pagebreak
% http://arxiv.org/pdf/quant-ph/0407159.pdf

%\blankpage                            % blank page with no running heads

%\printindex{author}{Author Index}     % to print author index

\end{document}